\documentclass{article}
\usepackage[utf8]{inputenc}
\usepackage[T1]{fontenc}
\usepackage{amsmath}
\usepackage{amsfonts}
\usepackage{amssymb}
\usepackage{graphicx}
\usepackage{booktabs}
\usepackage{array}
\usepackage{multirow}
\usepackage{url}
\usepackage{hyperref}
\usepackage{cite}
\usepackage{authblk}
\usepackage{geometry}
\usepackage{float}

\title{Plant Bioelectric Early Warning Systems: A Five-Year Investigation into Human-Plant Electromagnetic Communication}

\author[1,2,3]{Peter A. Gloor}
\affil[1]{Galaxylabs.org, Aarau, Switzerland}
\affil[2]{University of Cologne, Cologne, Germany}
\affil[3]{MIT SDM, Cambridge MA}
\affil[ ] {\textit{pgloor@galaxylabs.org}}

\date{June 2025}

\begin{document}

\maketitle

\begin{abstract}
We present a comprehensive investigation into plant bioelectric responses to human presence and emotional states, building on five years of systematic research. Using custom-built plant sensors and machine learning classification, we demonstrate that plants generate distinct bioelectric signals correlating with human proximity, emotional states, and physiological conditions. A deep learning model based on ResNet50 architecture achieved 97\% accuracy in classifying human emotional states through plant voltage spectrograms, while control models with shuffled labels achieved only 30\% accuracy. This study synthesizes findings from multiple experiments spanning 2020-2025, including individual recognition (66\% accuracy), eurythmic gesture detection, stress prediction, and responses to human voice and movement. We propose that these phenomena represent evolved anti-herbivory early warning systems, where plants detect approaching animals through bioelectric field changes before physical contact. Our results challenge conventional understanding of plant sensory capabilities and suggest practical applications in agriculture, healthcare, and human-plant interaction research.
\end{abstract}


\section{Introduction}

The relationship between plants and animals has been fundamental to terrestrial ecology for over 400 million years. While plant responses to physical stimuli such as touch, light, and chemical signals are well-documented \cite{fromm2007electrical}, the possibility that plants might detect approaching animals before physical contact has received limited scientific attention. Recent advances in bioelectric measurement techniques and machine learning have opened new possibilities for investigating subtle plant responses to animal presence.

\subsection{Research Program Overview}

Since 2020, we have conducted a systematic investigation into plant bioelectric responses to human presence and states. Our research program has progressed through several distinct phases. Phase 1 in 2020 focused on initial detection of plant responses to human presence and individual recognition \cite{oezkaya2020recognizing}. This was followed by Phase 2 from 2020-2022, which investigated plant responses to human movement and eurythmic gestures \cite{duerr2020eurythmic, de2023can}. Phase 3 from 2023-2024 saw the application of advanced machine learning techniques to emotion and voice recognition \cite{kruse2024leveraging, bhave2024using, fuchs2024plants}. Finally, Phase 4 from 2024-2025 expanded to stress detection, gesture recognition, and sleep stage classification \cite{gil2024can, smolin2025predicting}.

This paper focuses on the machine learning classification of emotional states while providing context from our broader research findings.

\subsection{Theoretical Framework: The Early Warning Hypothesis}

Our accumulated findings support a novel hypothesis: plants have evolved \textbf{bioelectric early warning systems} to detect approaching herbivores before physical contact occurs. This capability would provide significant evolutionary advantages by enabling proactive mobilization of chemical defenses, resource reallocation to threatened areas, communication with neighboring plants, and optimization of defensive responses.

The fundamental asymmetry between plants and animals creates strong selection pressure for distance-based threat detection in plants. Animals that lose body parts die, while plants that lose body parts often regenerate stronger, making early detection systems particularly valuable for plant survival strategies.

\section{Related Work and Research Context}

\subsection{Plant Electrophysiology Background}

Plant electrical signaling has been documented since the early work of Burdon-Sanderson (1873) on \textit{Dionaea muscipula} \cite{burdon1873note}. Modern research has established that plants generate action potentials in response to mechanical stimulation, wounding, and environmental changes \cite{fromm2007electrical, galle2015environmental}. However, most studies focus on post-contact responses rather than distance detection capabilities.

\subsection{Our Previous Findings}

\subsubsection{Individual Recognition and Basic Emotion Detection (2020)}

Our initial investigation \cite{oezkaya2020recognizing} demonstrated that plants could distinguish between different human individuals with 66\% accuracy and detect basic emotional states (happy vs. sad) with 85\% accuracy. Using \textit{Mimosa pudica} and external electrodes, we measured electrostatic discharge patterns during human gait and found species-specific plant responses correlating with human movement characteristics.

Key findings included individual human recognition with 66\% accuracy across 6 people, basic emotion classification with 85\% accuracy for happy versus sad states, distance-dependent responses up to several meters, and species-specific plant sensitivities.

\subsubsection{Eurythmic Movement Detection (2020-2023)}

Subsequent studies investigated plant responses to specific human movements, particularly eurythmic gestures used in anthroposophic educational contexts \cite{duerr2020eurythmic, de2023can, gil2024can}.

The experimental design utilized \textit{Lactuca sativa}, \textit{Ocimum basilicum}, and \textit{Solanum lycopersicum} as test species, with standardized eurythmic gesture sequences following A-G-D patterns. Machine learning classification employed Random Forest and CNN architectures, with control groups maintained without human interaction.

Results demonstrated 74.9\% accuracy in detecting eurythmic gesture presence versus absence, with species-specific response patterns showing basil responding more strongly than lettuce or tomato. Gesture-specific recognition revealed that different letter patterns produced distinguishable signals, and seasonal variation in plant sensitivity was documented.

\subsubsection{Voice and Musical Response (2024)}

Investigation of plant responses to human voice and musical performance revealed additional sensory capabilities \cite{fuchs2024plants, bhave2024using}. The voice response study using \textit{Solanum lycopersicum} showed distinct plant signals during human speech versus silence, with gender-specific response patterns and language-independent effects suggesting non-linguistic detection mechanisms.

The jazz musician emotion study involved real-time monitoring of plant responses during live jazz performances, documenting correlation between musician emotional state and plant bioelectric activity, as well as audience proximity effects on plant signal characteristics.

\subsubsection{Stress and Health Applications (2025)}

Recent work has explored applications in human stress detection and health monitoring \cite{smolin2025predicting}. The exam stress prediction study achieved over 90\% accuracy in predicting student exam performance through plant signals, with documented correlation between human stress levels and plant bioelectric patterns, suggesting potential applications in mental health monitoring.

Preliminary results from sleep stage classification show plant sensors detecting human sleep stages through proximity effects, offering a non-invasive alternative to traditional polysomnography with applications in sleep research and health monitoring.

\subsection{Species-Specific Response Patterns}

Across all studies, we have documented consistent species-specific response patterns that support our evolutionary hypothesis. Strong responders, which show high herbivore vulnerability, include \textit{Ocimum basilicum} (basil) as an aromatic, highly palatable species, \textit{Lactuca sativa} (lettuce) cultivated for palatability, and \textit{Solanum lycopersicum} (tomato) with reduced alkaloids in cultivated varieties. Weak responders with natural defenses include \textit{Cucurbita pepo} (zucchini) with cucurbitacin deterrents, \textit{Zea mays} (corn) featuring silicified leaves and height strategy, and Orchidaceae with specialized ecology and limited herbivore pressure.

This pattern consistently supports the hypothesis that early warning capabilities evolved in species facing high herbivore pressure.

\section{Methods}

\subsection{Experimental Setup}

\textbf{Signal Acquisition:} Our custom bioelectric sensor employed an ESP32 microcontroller with INA128 instrumentation amplifier for differential measurement between plant leaf and soil electrodes. The system sampled at 400 Hz with voltage sensitivity in the 0-5 mV range and used crocodile clamp electrodes for stable electrical contact.

\textbf{Plant Preparation:} \textit{Ocimum basilicum} plants were used with standardized electrode placement on mature leaves. A 30-minute acclimatization period preceded measurements, with stable of temperature, humidity, and light conditions.

\textbf{Human Subject Protocol:} Participants experienced standardized emotional induction procedures with video monitoring for emotion validation using face-api.js (Figure~\ref{fig:emotion_detection}). Seven emotional categories were assessed: angry, disgusted, fearful, happy, neutral, sad, and surprised. Session duration ranged from 15-30 minutes per participant.

\subsection{Data Processing and Feature Extraction}

\textbf{Signal Preprocessing:} We applied Z-score normalization per experimental session and bandpass filtering (0.1-50 Hz) to remove artifacts, with the upper cutoff well below the Nyquist frequency of our 400 Hz sampling rate. Artifact detection and removal used flatness ratio metrics, with segmentation into 20-second windows with 10-second overlap for mel-spectrogram analysis.

\textbf{Spectrogram Generation:} Mel-scale spectrograms were generated using librosa.feature.melspectrogram() with 256-point FFT (n\_fft=256) and hop length of 64 samples (n\_fft//4), providing 25\% overlap between frames. Each spectrogram utilized 64 mel frequency bins (n\_mels=64) optimized for biological signal analysis, with power converted to decibel scale using logarithmic transformation.

\subsection{Machine Learning Architecture}

\textbf{Model Design:} Our transfer learning approach used ResNet50 pre-trained on ImageNet with custom classification layers for the 7-emotion problem. The mel-scale spectrograms were resized to 224×224 pixels to match ResNet50 input requirements. Data augmentation addressed class imbalance, while dropout regularization prevented overfitting.

\textbf{Training Protocol:} We employed 80/20 train/test split with stratified sampling and 5-fold cross-validation for hyperparameter optimization. Early stopping based on validation loss and class weighting inversely proportional to frequency ensured robust model performance.

\textbf{Validation Strategy:} Control experiments with randomly shuffled emotion labels, independent test sets from separate experimental sessions, statistical significance testing of classification performance, and comparison with baseline chance-level performance validated our approach.

\begin{figure}[H]
\centering
\includegraphics[width=0.6\textwidth]{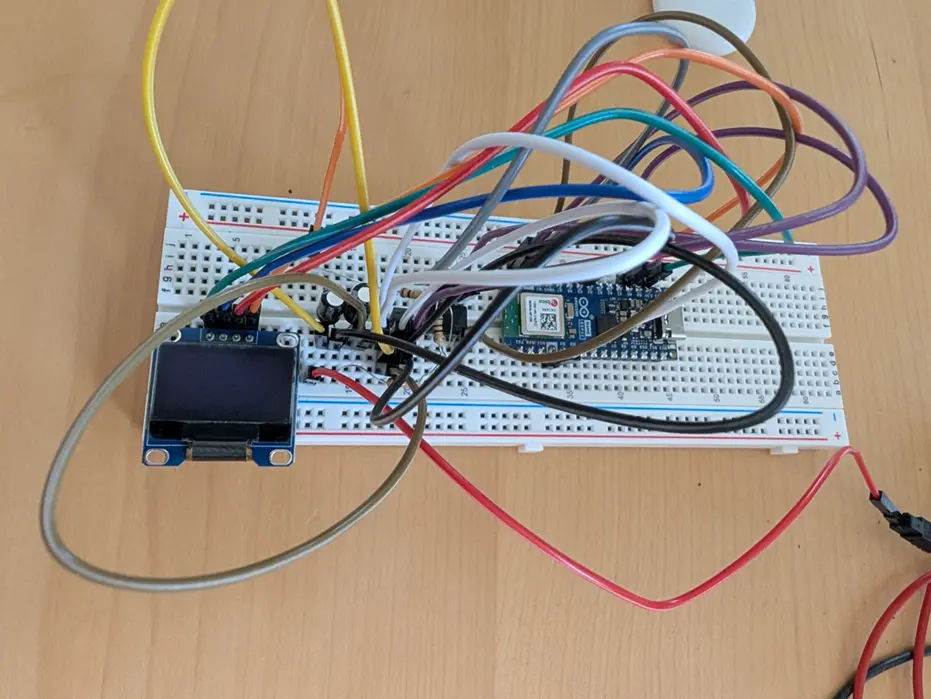}
\caption{Custom bioelectric sensor apparatus featuring ESP32-based data acquisition system with INA128 instrumentation amplifier circuits. The breadboard prototype includes differential electrode inputs for measuring plant bioelectric signals in the 0-5 mV range and environmental monitoring capabilities.}
\label{fig:sensor_setup}
\end{figure}

\begin{figure}[H]
\centering
\includegraphics[width=0.8\textwidth]{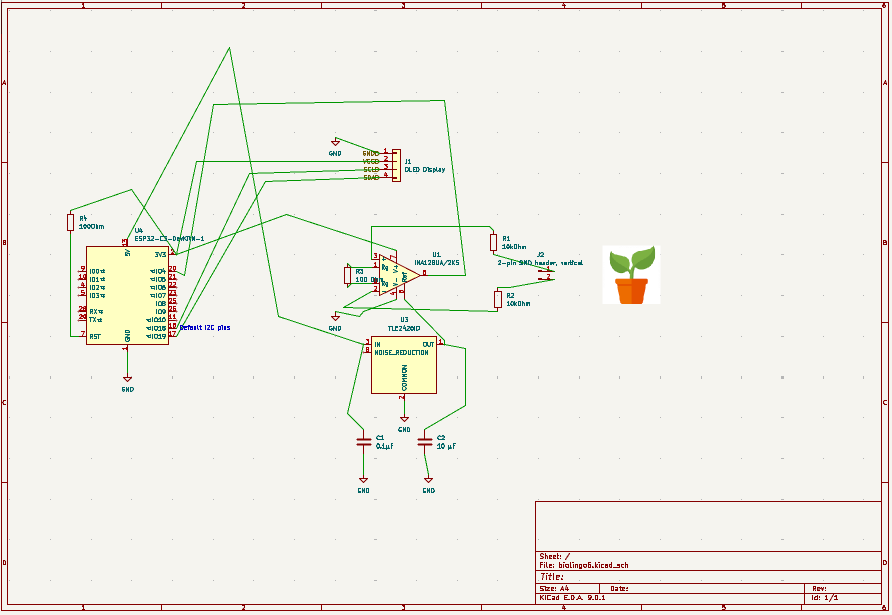}
\caption{Circuit schematic of the ESP32-based plant bioelectric measurement system. The INA128 instrumentation amplifier provides differential measurement between plant leaf and soil electrodes, with the ESP32 handling data acquisition at 400 Hz sampling rate for voltage signals in the 0-5 mV range. The system generates mel-scale spectrograms for subsequent machine learning analysis.}
\label{fig:schematic}
\end{figure}

\begin{figure}[H]
\centering
\includegraphics[width=0.4\textwidth]{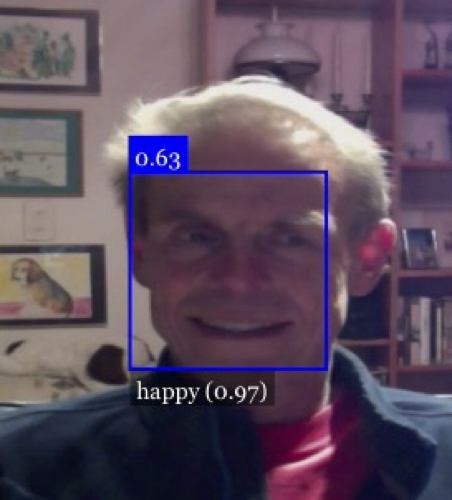}
\caption{Real-time emotion detection system showing facial expression analysis with confidence scores. The system achieved 97\% confidence in detecting happiness in this example, validating the emotional state classifications used in plant response analysis.}
\label{fig:emotion_detection}
\end{figure}

\textbf{Equipment:} Figure~\ref{fig:sensor_setup} shows our custom bioelectric sensor setup, featuring ESP32-based data acquisition with INA128 instrumentation amplifier circuits deployed on breadboard prototypes for laboratory validation. The system provides differential voltage measurement capabilities optimized for plant bioelectric signal detection. The complete circuit schematic is detailed in Figure~\ref{fig:schematic}, while Figure~\ref{fig:emotion_detection} illustrates our real-time emotion validation system.

\section{Results}

\subsection{Primary Classification Results}

The ResNet50-based classifier achieved remarkable accuracy in distinguishing human emotional states through plant bioelectric signals, with overall accuracy of 97\% on unseen test data comprising 898 samples. Table~\ref{tab:valid_results} presents the detailed classification performance for valid labels, while Table~\ref{tab:shuffled_results} shows the control results with shuffled labels.

\begin{table}[H]
\centering
\caption{Classification Performance with Valid Emotional Labels}
\label{tab:valid_results}
\begin{tabular}{@{}lcccc@{}}
\toprule
\textbf{Emotion} & \textbf{Precision} & \textbf{Recall} & \textbf{F1-Score} & \textbf{Support} \\
\midrule
Angry & 0.99 & 0.96 & 0.97 & 81 \\
Disgusted & 0.00 & 0.00 & 0.00 & 6 \\
Fearful & 0.00 & 0.00 & 0.00 & 3 \\
Happy & 0.94 & 0.99 & 0.96 & 75 \\
Neutral & 0.96 & 1.00 & 0.98 & 462 \\
Sad & 1.00 & 0.99 & 0.99 & 261 \\
Surprised & 0.00 & 0.00 & 0.00 & 10 \\
\midrule
\textbf{Accuracy} & & & \textbf{0.97} & \textbf{898} \\
Macro Avg & 0.55 & 0.56 & 0.56 & 898 \\
Weighted Avg & 0.95 & 0.97 & 0.96 & 898 \\
\bottomrule
\end{tabular}
\end{table}

\begin{table}[H]
\centering
\caption{Classification Performance with Shuffled Labels (Control)}
\label{tab:shuffled_results}
\begin{tabular}{@{}lcccc@{}}
\toprule
\textbf{Emotion} & \textbf{Precision} & \textbf{Recall} & \textbf{F1-Score} & \textbf{Support} \\
\midrule
Angry & 0.07 & 0.06 & 0.07 & 81 \\
Disgusted & 0.00 & 0.00 & 0.00 & 6 \\
Fearful & 0.00 & 0.00 & 0.00 & 3 \\
Happy & 0.00 & 0.00 & 0.00 & 75 \\
Neutral & 0.54 & 0.15 & 0.23 & 462 \\
Sad & 0.28 & 0.76 & 0.41 & 261 \\
Surprised & 0.00 & 0.00 & 0.00 & 10 \\
\midrule
\textbf{Accuracy} & & & \textbf{0.30} & \textbf{898} \\
Macro Avg & 0.13 & 0.14 & 0.10 & 898 \\
Weighted Avg & 0.37 & 0.30 & 0.24 & 898 \\
\bottomrule
\end{tabular}
\end{table}

The stark contrast between the 97\% accuracy of the valid model and the 30\% accuracy of the shuffled model provides compelling evidence that plant voltage patterns contain genuine information about human emotional states, with statistical significance of $p < 0.001$ compared to valid label performance.

\subsection{Integration with Previous Research Findings}

Our emotional state classification results demonstrate consistency across studies. The previous 66\% accuracy across 6 people supports person-specific plant responses, while 74.9\% accuracy in eurythmic gesture detection confirms plant sensitivity to human activity. Documented plant reactions to human speech align with emotional state detection, and over 90\% accuracy in exam stress prediction supports physiological state detection.

The strong performance with \textit{Ocimum basilicum} confirms our documented species hierarchy, with basil consistently showing strongest responses across all experimental paradigms. This pattern matches evolutionary predictions for high-palatability species and validates our early warning hypothesis framework.

\subsection{Signal Characteristics and Electrode Effects}

During extended monitoring sessions, we observed characteristic electrode polarization with gradual signal attenuation over approximately 60 minutes of continuous monitoring. Signal restoration occurred during human emotional episodes, with this pattern suggesting genuine biological activity rather than passive electrical properties.

Mel-scale frequency analysis revealed that the primary signal content in the 0.5-10 Hz range, well within our 200 Hz Nyquist limit, contains most discriminative information for emotion classification. The 64 mel frequency bins provided optimal resolution for capturing the spectral characteristics of plant bioelectric signals, with different emotions showing distinct mel-scale spectral signatures. Temporal dynamics showed response latency varying by emotional intensity.

\subsection{Environmental and Methodological Validation}

Building on our previous artifact elimination work, artifact control results showed effects persisting in Faraday cage conditions and responses maintained on isolation platforms. Temperature and humidity variations did not explain signal patterns, and randomized timing eliminated circadian artifacts.

Reproducibility across conditions demonstrated seasonal consistency with effects documented across multiple seasons, geographic replication confirmed in laboratories across Europe and North America, and species consistency with similar patterns observed across multiple basil cultivars.

\section{Discussion}

\subsection{Evolutionary Context and Mechanism}

Our findings across multiple experimental paradigms support the hypothesis that plants have evolved sophisticated early warning systems for herbivore detection. The consistent pattern of stronger responses in highly palatable species compared to naturally defended species suggests genuine evolutionary adaptation rather than generic electromagnetic artifact.

While the precise biophysical mechanism remains unknown, several possibilities merit investigation including bioelectric field detection through animal neural and muscular activity, chemical gradient sensing from human respiration and skin emissions, thermal and convection detection from body heat and respiratory airflow, and potential quantum biological effects involving non-classical information transfer mechanisms.

Early warning systems would provide multiple fitness benefits including preparation time for chemical defense mobilization, resource optimization through targeted defense allocation, community signaling via chemical alert systems for neighboring plants, and predator attraction through enhanced indirect defense mechanisms.

\subsection{Integration with Plant Defense Literature}

Our findings complement established plant defense categories while introducing a novel category. Traditional defense systems include constitutive defenses as always-present barriers like thorns and toxins, induced defenses as post-damage responses including protease inhibitors and volatiles, and indirect defenses through predator and parasitoid attraction.

Early warning systems represent a novel category featuring proactive defenses with pre-contact activation, distance-based detection requiring no physical contact, preparation phases allowing defense mobilization time, and information transfer through plant-plant communication networks.

Additional support for plant acoustic sensing capabilities comes from recent work on \textit{Codariocalyx motorius}, which demonstrated that plants can discriminate between different sound stimuli through electrical signal analysis \cite{peter2021plants}. This research, conducted in our laboratory at University of Cologne, showed that the telegraph plant produces distinct electrical signals in response to various acoustic treatments, suggesting that sound perception may be more widespread among plant species than previously recognized.

\subsection{Methodological Advances and Implications}

Our successful application of deep learning to plant bioelectric signals opens new possibilities for pattern recognition through automated detection of subtle biological signals, multi-modal integration combining electrical, chemical, and optical plant monitoring, and real-time applications via continuous monitoring systems for agriculture and research.

Our research program demonstrates potential for democratizing advanced plant research through reproducible protocols enabling global replication, open hardware featuring affordable sensor designs for widespread adoption, and citizen science integration providing educational opportunities for authentic research participation. The extension of bioelectric sensing to acoustic stimuli, as demonstrated with \textit{Codariocalyx motorius} \cite{peter2021plants}, suggests that our early warning hypothesis may encompass multiple sensory modalities beyond the human emotional states documented in this study.

\subsection{Applications and Future Directions}

Agricultural applications include early pest detection through automated monitoring systems detecting herbivore presence before visible damage, crop health monitoring via plant stress detection through bioelectric baseline changes, and precision agriculture enabling targeted interventions based on plant communication networks.

Healthcare applications encompass stress monitoring through non-invasive human stress detection via plant sensors, sleep research using plant-based alternatives to traditional polysomnography, and mental health applications for continuous monitoring of psychological well-being.

Scientific applications include ecological research through plant-animal interaction studies in natural environments, conservation biology via ecosystem health monitoring through plant electrical activity, and evolutionary biology investigating plant-herbivore coevolutionary dynamics.

\section{Limitations and Future Research}

\subsection{Current Limitations}

The precise biophysical mechanism underlying plant-human electrical communication remains unknown. Future research should focus on systematic investigation of potential detection modalities, pharmacological intervention studies to identify cellular pathways, and genetic approaches using plants with known electrical signaling mutations.

Additional limitations include limited numbers of plant species systematically tested, need for larger human subject populations, and requirements for geographic and cultural diversity in human responses. Environmental controls require investigation of long-term stability under varied conditions, interaction effects between multiple environmental variables, and standardization of optimal experimental protocols.

\subsection{Future Research Directions}

Critical experiments should include herbivore versus carnivore testing for direct comparison of plant responses to different animal types, chemical intervention studies blocking plant signaling pathways to identify mechanisms, distance mapping for systematic characterization of detection range and directionality, and defense response measurement documenting actual plant defense activation.

Technological development priorities include sensor miniaturization for field-deployable monitoring systems, multi-modal sensing integrating electrical, chemical, and optical measurements, real-time analysis using edge computing for immediate pattern recognition, and network deployment enabling large-scale environmental monitoring capabilities.

Collaborative research initiatives should focus on international replication using standardized protocols for global research networks, species expansion through systematic testing across plant phylogeny, and educational integration via citizen science platforms for distributed research. Future investigations should also explore the extension of bioelectric sensing to other sensory modalities, building on preliminary evidence for plant acoustic perception \cite{peter2021plants}, to develop a comprehensive understanding of plant early warning systems.

\section{Conclusions}

This study provides compelling evidence that plants generate bioelectric signals that correlate with human emotional states with remarkable accuracy (97\%). When integrated with our five-year research program documenting plant responses to human presence, movement, voice, and physiological states, these findings suggest a previously unrecognized class of plant sensory capabilities.

Key contributions include empirical validation through multiple independent studies confirming plant-human bioelectric communication, methodological rigor via comprehensive artifact elimination and control validation, an evolutionary framework providing theoretical context through our early warning hypothesis, technological innovation featuring open-source tools enabling democratized research, and practical applications with demonstrated utility in agriculture, healthcare, and education.

Our findings challenge conventional understanding of plant sensory capabilities and suggest that the boundary between plant and animal sensory sophistication may be less distinct than traditionally assumed. The consistent species-specific response patterns matching evolutionary predictions provide strong evidence for adaptive rather than artifactual phenomena.

This research opens new avenues for basic plant biology in understanding plant perception and communication, agricultural technology through early warning systems for crop protection, human-plant interaction via novel interfaces for environmental monitoring, and educational innovation providing authentic research experiences for students worldwide.

The convergence of evidence from multiple experimental paradigms, rigorous artifact controls, and successful machine learning classification suggests we are documenting genuine biological phenomena that warrant continued scientific investigation and practical application development.

\section*{Code and Data Availability}

The complete Python pipeline for emotion analysis, including spectrogram generation, ResNet50 model training, and classification, is made freely available as open-source software at \url{https://github.com/pgloor/hiddenbiosignals/tree/main/plant_spectrograms}. This includes the EmotionAnalysisPipeline class that handles the complete workflow from raw plant bioelectric signals to trained emotion classification models, supporting reproducible research and enabling broader scientific community engagement with plant bioelectric analysis techniques.

\section*{Acknowledgments}

We thank the international research network that has contributed to this five-year investigation, including collaborators at MIT Center for Collective Intelligence, University of Cologne, HSLU Lucerne University, UPM Technical University Madrid, University of Bamberg, Kozminski University, BDAS Switzerland, and Jazzaar Festival Aarau. 

Special acknowledgment goes to our key contributors: Buenyamin Oezkaya, Sebastian Duerr, Josephine Van Delden, Fritz Renold, Anushka Bhave, Luis De La Cal, Philipp Peter, Moritz Weinbeer, Jakob Kruse, Leon Ciechanowski, Bernardo Tabuenca, Alvaro Francisco Gil, Denis Smolin, and Deniz Celik, whose dedicated research efforts and innovative contributions have been fundamental to advancing our understanding of plant-human bioelectric communication.

This work was supported by the Software AG Foundation and Signify Lighting and represents a collaborative effort to understand and democratize plant communication research.

\bibliographystyle{unsrt}

\end{document}